\documentclass[superscriptaddress, nofootinbib, amsmath,amssymb, aps, prd, twocolumn]{revtex4-2}

\usepackage{comment}
\usepackage{graphicx}
\usepackage{dcolumn}
\usepackage{bm}
\usepackage{hyperref}
\usepackage{xcolor}
\usepackage{acronym}
\usepackage{lipsum}

\usepackage[normalem]{ulem}

\newacro{adm}[ADM]{Arnowitt-Deser-Misner}
\newacro{amr}[AMR]{adaptive mesh refinement}
\newacro{bbh}[BBH]{binary black hole}
\newacroplural{bbh}[BBHs]{binary black holes}
\newacro{bh}[BH]{black hole}
\newacroplural{bh}[BHs]{black holes}
\newacro{bhns}[BHNS]{black hole-neutron star}
\newacro{bhpt}[BHPT]{black hole perturbation theory}
\newacro{bns}[BNS]{binary neutron star}
\newacro{bf}[BF]{Bayes' factor}
\newacro{cbc}[CBC]{compact binary coalescence}
\newacro{ce}[CE]{Cosmic Explorer}
\newacro{cce}[CCE]{Cauchy characteristic extraction}
\newacro{da}[DA]{data analysis}
\newacro{et}[ET]{Einstein Telescope}
\newacro{efe}[EFE]{Einstein Field Equation}
\newacro{eob}[EOB]{Effective-One-Body}
\newacro{eom}[EOM]{equations of motion}
\newacro{fd}[FD]{frequency domain}
\newacro{fre}[FRE]{finite-radius extraction}
\newacro{fd}[FD]{finite differencing}
\newacro{fft}[FFT]{Fast Fourier transform}
\newacro{gw}[GW]{gravitational wave}
\newacro{gr}[GR]{general relativity}
\newacro{grb}[GRB]{gamma-ray burst}
\newacro{grhd}[GRHD]{general-relativistic hydrodynamics}
\newacro{gwosc}[GWOSC]{Gravitational Wave Open Science Center}
\newacro{gwtc1}[GWTC-1]{the first gravitational-wave transients catalog}
\newacro{gsf}[GSF]{Gravitational Self Force}
\newacro{hm}[HM]{Higher mode}
\newacroplural{hm}[HMs]{Higher modes}
\newacro{hpc}[HPC]{high-performance computing}
\newacro{ifo}[IFO]{interferometer}
\newacro{imr}[IMR]{inspiral-merger-ringdown}
\newacro{im}[IMR]{inspiral-to-merger}
\newacro{kagra}[KAGRA]{Kamioka Gravitational Wave Detector}
\newacro{ligo}[LIGO]{Laser Interferometer Gravitational-Wave Observatory}
\newacro{lisa}[LISA]{Laser Interferometer Space Antenna}
\newacro{lr}[LR]{Light Ring}
\newacro{lso}[LSO]{Last Stable Orbit}
\newacro{lvc}[LVC]{LIGO-Virgo Collaboration}
\newacro{lvk}[LVK]{LIGO-Virgo-Kagra}
\newacro{lo}[LO]{leading order}
\newacro{nn}[NN]{neural network}
\newacro{ns}[NS]{neutron star}
\newacroplural{ns}[NSs]{neutron stars}
\newacro{nr}[NR]{numerical relativity}
\newacro{nqc}[NQC]{next-to-quasicircular}
\newacro{nlo}[NLO]{next-to-leading order}
\newacro{nnlo}[NNLO]{next-to-next-to-leading order}
\newacro{n3lo}[N3LO]{next-to-next-to-next-to-leading order}
\newacro{n4lo}[N3LO]{next-to-next-to-next-to-next-to-leading order}
\newacro{ode}[ODE]{Ordinary Differential Equation}
\newacroplural{ode}[ODEs]{Ordinary Differential Equations}
\newacro{pn}[PN]{post-Newtonian}
\newacro{pm}[PM]{post-Minkowskian}
\newacro{pe}[PE]{parameter estimation}
\newacro{psd}[PSD]{power spectral density}
\newacroplural{psd}[PSD]{power spectral densities}
\newacro{pa}[PA]{post-adiabatic}
\newacro{pinn}[PINN]{physics-informed neural network}
\newacro{qnm}[QNM]{quasi-normal mode}
\newacro{qc}[QC]{quasi-circular}
\newacro{snr}[SNR]{signal-to-noise ratio}
\newacro{spa}[SPA]{stationary-phase approximation}
\newacro{sxs}[SXS]{Simulating eXtreme Spacetimes}
\newacro{td}[TD]{time domain}
\newacro{ng}[NG]{Nect Generation}
\newacro{dcs}[dCS]{dynamical Chern-Simons}
\newacro{bgr}[BGR]{beyond-\ac{gr}}
\newacro{spec}[SpEC]{Spectral Einstein Code}
\newacro{edgb}[EdGB]{Einstein-dilaton-Gauss-Bonnet}
\newacro{emri}[EMRI]{extreme mass-ratio inspiral}
\newacro{gwtc}[GWTC]{Gravitational-Wave Transient Catalog}
\newacro{far}[FAR]{false alarm rate}
\newacro{ci}[CI]{credible interval}

\begin{document}

\title{\textbf{Puncture Evolution with Physics-Informed Neural Networks}}

\author{Estuti Shukla}
\email{eps5617@psu.edu}
\affiliation{Institute for Gravitation and the Cosmos, The Pennsylvania State University, University Park, PA 16802}
\affiliation{Department of Physics, The Pennsylvania State University, University Park, PA 16802}
\author{Ashwin Suriyanarayanan}
\affiliation{School of Mechanical Engineering, Purdue University, West Lafayette IN 47907, USA}
\author{David Radice}
\affiliation{Institute for Gravitation and the Cosmos, The Pennsylvania State University, University Park, PA 16802}
\affiliation{Department of Physics, The Pennsylvania State University, University Park, PA 16802}
\affiliation{Department of Astronomy \& Astrophysics, The Pennsylvania State University, University Park PA 16802, USA}
\author{Romit Maulik}
\affiliation{School of Mechanical Engineering, Purdue University, West Lafayette IN 47907, USA}
\author{Sebastiano Bernuzzi}
\affiliation{Theoretisch-Physikalisches Institut, Friedrich-Schiller-Universit\"at Jena, 07743, Jena, Germany}

\begin{abstract}
We investigate the application of \acp{pinn} to solve the vacuum Einstein equations in spherical symmetry. Our neural network architecture is designed to solve the $1+1$D moving-puncture evolution, by minimizing a loss that combines a data term, the Ricci tensor of the reconstructed spacetime, and the moving-puncture gauge conditions. It represents the lapse, the shift and the Z4c conformal variables \(\chi\) and \(\tilde\gamma_{rr}\), which remain regular at the puncture. We consider the evolution of a single, non-spinning black-hole in the puncture gauge and show that \acp{pinn} can interpolate \ac{nr} data, predict its evolution, and reconstruct spacetime in regions where no data is provided. As an interpolant, the network provides a compact, differentiable representation of the \ac{nr} data. Given only early-time and outer-boundary data, it evolves the puncture forward in time and reproduces the main features of the \ac{nr} solution. Given only exterior data and no gauge condition, an ill-posed inverse problem, it recovers the interior of a Schwarzschild spacetime in coordinates of its own choosing, which we verify by learning the coordinate transformation to the Kerr--Schild form. Finally, we assess the strengths and limitations of \acp{pinn} for \ac{nr} and outline their prospects for future applications.

\end{abstract}
\maketitle

\acrodef{PINN}{Physics-Informed Neural Network}

\section{Introduction}
The Einstein field equations of general relativity are a set of ten nonlinear partial differential equations of mixed hyperbolic and elliptic character, prescribing the geometry of spacetime in general relativity. Due to their
complex nature, exact solutions are known only in a handful of cases, describing spacetimes with a high degree of symmetry \cite{Muller:2009bw}. Perturbative and
approximate solutions are also known \cite{Chandrasekhar:1984siy}. However, \acf{nr} is the only tool able to solve the equations of
general relativity for \emph{generic} spacetimes to arbitrary accuracy~\cite{2010nure.book.....B}.

Most \ac{nr} calculations are either concerned with the
construction of stationary solutions, or solve the Cauchy (or characteristic) problem of general relativity. In the latter case, initial data is specified on a spacelike (or null)
hypersurface, then, its future development is computed, typically by advancing it over a set of hypersurfaces to form a foliation of (a portion of) the spacetime of interest \cite{Gourgoulhon:2007ue,Bruhatbook}. For spacelike foliations, this approach is known as $3+1$ \cite{Darmois1927,Lichnerowicz1944, FouresBruhat1952, FouresBruhat1956} or the \ac{adm} decomposition
\cite{Arnowitt:1959ah}.
In the $3+1$ formalism, Einstein equations cleanly split into a set of 4 elliptic equations (the constraints) and 6 hyperbolic (or evolution) equations. It can be shown that, if the initial data is chosen to satisfy the constraints, then its evolution will also preserve the constraints. So the standard approach in \ac{nr} is to separately solve for the constraints, at the time when initial data is constructed, and then use the hyperbolic equations to propagate the data forward in ``time''.

While the $3+1$ formalism was established by 1950s, stable numerical solution of the Cauchy problem of general relativity required substantial new developments, since it was later discovered that the \ac{adm}
equations, together with commonly used gauge choices, are only weakly hyperbolic \cite{PhysRevD.70.044031,2008itnrbook}. Modern
\ac{nr} evolution codes employ either the generalized-harmonic (GH) formalism \cite{Pretorius:2004jg, Lindblom:2005qh}, the Baumgarte, Shapiro,
Shibata, Nakamura (BSSN) formalism \cite{Shibata:1995we, Baumgarte:1998te}, or
variants of the Z4 formalism \cite{Bona:2003fj}, such as Z4c
\cite{Bernuzzi:2009ex,Hilditch_2013} and CCZ4 \cite{Alic:2011gg}. The resulting equations are
then solved using finite differencing, or spectral methods. These techniques have
been applied to a wide range of problems of theoretical physics and astrophysics
interest: from the collapse of rotating stars \cite{Stark:1985da,
Shapiro:1991zza, Shibata:2002br, Dimmelmeier:2002bk, Baiotti:2004wn, Ott:2006eu,
Sekiguchi:2010ep, Ott:2012kr, Mosta:2014jaa, Mosta:2015ucs, Kuroda:2016bjd}, to
compact binary mergers \cite{Pretorius:2005gq, Campanelli:2005dd, Baker:2005vv,
Campanelli:2007cga, Scheel:2008rj, Boyle:2019kee, Shibata:2006nm,
Baiotti:2008ra, Hotokezaka:2012ze, Kiuchi:2015sga, Dietrich:2017aum,
Radice:2017lry, Palenzuela:2021gdo, Mitman:2022qdl, Cheung:2022rbm,
Ferguson:2023vta, Kiuchi:2022ubj, Kiuchi:2023obe, Combi:2023yav, Most:2023sft,
Espino:2023dei, Rashti:2024yoc, Musolino:2024sju, Scheel:2025jct, Black:2026scc}
and gravitational scattering \cite{Shibata:2008rq, Sperhake:2009jz, Gold:2012tk,
Damour:2014afa, Hopper:2022rwo, Rettegno:2023ghr, Albanesi:2024xus}, among
others.

 \Acfp{pinn}\cite{Raissi:2017zsi,2019arXiv190704502L, 2021arXiv210709443Z,
Haghighat2021, Lu2021} are a new approach to solve partial differential equations
developed in the context of scientific machine learning. \Acp{pinn} represent
the solution of a partial differential equation as a neural-network (NN), whose
weights and biases are learned by minimizing a loss function which includes both
initial/boundary data and the residual of the partial differential equation (the
``physics'' loss term). These methods find justification in the universal approximation theorem \cite{Cybenko:1989iql, Hornik:1989yye, FUNAHASHI1989183,HORNIK1990551,  2024arXiv240712895A}, which ensures that the solution of any partial differential equation can be represented with a NN to arbitrarily high precision.
\acp{PINN} can be thought of as non-linear generalizations of Galerkin methods \cite{Lu2021}.

\Acp{PINN} have been successfully applied to a large variety of problems,
including fluid mechanics \cite{Mao2020, Raissi2020, Pavlou:2024aaj, Cheung:2025kda},
cosmology \cite{Thomsen:2025yfi}, differential geometry
\cite{Hashimoto:2025zmi}, and black-hole perturbation theory \cite{Luna:2022rql,
Patel:2024wzo, Cornell:2024azz, Pombo:2025urp}. In the context of numerical
relativity, Ref.~\cite{Cranganore:2025vsu} demonstrated the use of \acp{PINN} as
a way to obtain a differentiable representation of a (\ac{nr}
generated) spacetime, while Ref.~\cite{Zhou:2026haz} employed \acp{PINN} for the
construction of binary black hole (BH) initial data, that is for the solution of the
elliptic sector of Einstein equations. Ref.~\cite{Ferrer-Sanchez:2025cns} used \acp{PINN} to study the dynamics of the
collapse of scalar fields, extending the classical work by
Choptuik~\cite{Choptuik:1992jv} by solving the Cauchy problem in 1+1D (spherical
symmetry). Ref.~\cite{altgrpinn} combined \acp{pinn} with a pseudo-spectral refinement step to construct stationary rotating black-hole solutions in higher-curvature theories of gravity. Ref.~\cite{li2023solvingeinsteinequationsusing} used \acp{pinn} to recover the Schwarzschild and Reissner--Nordstr\"om metric, adopting a static ansatz in Schwarzschild coordinates. Additionally, Ref.~\cite{bayona2026} reproduced the Schwarzschild solution and the axisymmetric $q$-metric, with a focus on how boundary conditions and domain decomposition affect \ac{pinn} training. Most recently, Ref.~\cite{Carretero:2026vef} used \acp{PINN} to solve the Einstein equations coupled to a scalar field in the generalized harmonic gauge.

In this work, we explore the potential of using \acp{PINN} to solve the Einstein equations, in the puncture gauge, in the reduced settings of spherical symmetry. In contrast to \cite{li2023solvingeinsteinequationsusing}, that assumes a static metric ansatz, we do not impose stationarity. We consider, in addition to the solution of the ``standard''
Cauchy problem, the use of \acp{PINN} to solve \emph{inverse} problems, in which
data is not prescribed on a Cauchy hypersurface. We critically assess the
capability of these schemes and discuss possible future extensions and
applications. We remark that Ref.~\cite{SchettiniGherardini:2026bdb} has also
recently proposed a \ac{PINN} approach to construct BH spacetimes. Our
approaches are complementary: our goal is to use \acp{PINN} to reconstruct
spacetimes given incomplete data, while \cite{SchettiniGherardini:2026bdb} aims
to construct general spacetimes satisfying specified algebraic properties.

The rest of this article is organized as follows. In Sec.~\ref{sec:methods}, we
describe in detail the architecture we have developed. In
Sec.~\ref{sec:results}, we apply \acp{PINN} to study a Schwarzschild black-hole
spacetime in different gauges. Finally, we summarize our results and discuss
future work in Sec.~\ref{sec:conclusions}. We use geometrized units with $G = c = 1$. Greek indices ($\mu, \nu, \ldots = 0$--$3$) label components of four-dimensional spacetime tensors, while Latin indices ($i, j, \ldots = 1$--$3$) label components of spatial tensors on the $t = \text{const}$ hypersurfaces of the $3+1$ foliation. Repeated indices are summed over, following the Einstein summation convention.

\section{Methodology}
\label{sec:methods}

\subsection{Input}
\label{sec:methods.input}

In this work, we investigate spherically symmetric spacetime solution of the vacuum Einstein field equations. The network takes in the two-dimensional coordinate \((t,r)\) as input and predicts
the four spherically symmetric metric fields, 
\begin{equation}
    \mathcal{N}(t,r)
    =
    \left(\alpha(t,r),\,\beta^r(t,r),\,\chi(t,r),\,
    \tilde{\gamma}_{rr}(t,r)\right).
\end{equation}
Here, \(\alpha\) is the lapse function, \(\beta^r\) is the radial component of the
shift vector, \(\chi\) is the conformal factor, and
\(\tilde{\gamma}_{rr}\) is the radial component of the conformal spatial
metric.

In $3+1$ decomposition, the line element is given by,
\begin{equation}
\begin{split}
    ds^2 = g_{\mu\nu}dx^{\mu}&dx^{\nu} =  -\alpha^2 dt^2 \\
    &+ \gamma_{ij}\left(dx^i+\beta^i dt\right)
      \left(dx^j+\beta^j dt\right),
  \end{split}
\end{equation}
where \(\gamma_{ij}\) is the spatial metric. Following the Z4c conformal
decomposition \cite{PhysRevD.88.084057}, we write
\begin{equation}
    \gamma_{ij}=\chi^{-1}\tilde{\gamma}_{ij},
    \qquad
    \det\!\left(\tilde{\gamma}_{ij}\right)=\det\!\left(\hat{\gamma}_{ij}\right),
\end{equation}
where \(\hat{\gamma}_{ij}\) is the flat spatial metric in spherical
coordinates. 

For spherically symmetric spacetime, the four network outputs reconstruct the full spacetime line element as
\begin{align}
    ds^2 ={}&
    \left[-\alpha^2+
    (\beta^r)^2\frac{\tilde{\gamma}_{rr}}{\chi}\right]dt^2
    +2\beta^r\frac{\tilde{\gamma}_{rr}}{\chi}\,dt\,dr
    +\frac{\tilde{\gamma}_{rr}}{\chi}\,dr^2 \nonumber\\
    &+\frac{r^2}{\chi\sqrt{\tilde{\gamma}_{rr}}}
    \left(d\theta^2+\sin^2\theta\,d\phi^2\right). \label{eq:metric}
\end{align}

We rescale the physical coordinate $(t,r)$ before feeding them into the network by mapping them to $(\hat t,\hat r) \in[-1,1]^2$. This scaling is given by,
\begin{equation}
  \hat x = 2\,\frac{x - x_{\rm lo}}{x_{\rm hi}-x_{\rm lo}} - 1 ,
\end{equation}
where $x_{\rm lo}$ and $x_{\rm hi}$ are the lower and upper bounds of each coordinate over the training domain. The network outputs native variables
$(z_\alpha,z_{\beta^r},z_\chi,z_{\tilde{\gamma}_{rr}})$, from which the physical
fields are reconstructed as
$\alpha=e^{z_\alpha}$, $\beta^r=z_{\beta^r}$, $\chi=e^{z_\chi}$, and
$\tilde{\gamma}_{rr}=e^{z_{\tilde{\gamma}_{rr}}}$. The exponential parametrization enforces positivity of the lapse, conformal factor,
and radial conformal metric component. As a consequence, the physical metric and its inverse which contains $1/\alpha^2$, $1/\chi$ and $1/\sqrt{\tilde{\gamma}_{rr}}$  stay finite throughout the training. We found this necessary to ensure a stable training.

The domain of $(t,r)$ depends on the application considered in Sec.~\ref{sec:results}. For the physics-informed interpolation (Sec.~\ref{sec:results.interp}) and the Cauchy problem (Sec.~\ref{sec:results.cauchy}), we use the \ac{nr} data of a puncture evolution, which covers $r \in [0.505, 170]$. In the interpolation problem the network is trained over the whole numerical domain, $t \in [0, 100]$, with data provided everywhere. In the Cauchy problem the domain is restricted to $t \in [0, 60]$, and data is provided only for $t \le 10$ (initial data) and $r \ge 150$ (outer boundary). For the boundary value problem (Sec.~\ref{sec:results.bval}), the domain is $t \in [0, 50]$ and $r \in [2, 20]$, with analytical Kerr-Schild data provided only for $r \ge 15$.

\subsection{Loss function}
The total training objective is given by,
\begin{equation}
\begin{split}
\mathcal{L}
=
w_{\rm data}\mathcal{L}_{\rm data}
& +
w_{\rm phy}\mathcal{L}_{\rm phy} \\
& +
w_{\rm lapse}\mathcal{L}_{\rm lapse}
+
w_{\rm shift}\mathcal{L}_{\rm shift}.
\label{eq:loss}
\end{split}
\end{equation}
where $w_{\rm data}$, $w_{\rm phy}$, $w_{\rm lapse}$ and $w_{\rm shift}$ are the weights of the various loss components. For simplicity, we keep these weights constant throughout the training. The supervised data loss term is evaluated only on the available labeled points,
\begin{equation}
\mathcal{L}_{\rm data}
=
\frac{1}{N_{\rm data}}
\sum_{i=1}^{N_{\rm data}}
\left[
\mathbf{u}_{\theta}(t_i,r_i)
-
\mathbf{u}^{\rm NR}(t_i,r_i)
\right]^2.
\end{equation}
where $\mathbf{u}_\theta=(z_\alpha,z_{\beta^r},z_\chi,z_{\tilde{\gamma}_{rr}})$.

The geometric physics loss $\mathcal{L}_{\rm phy}$ is obtained by reconstructing the spacetime metric
from the network outputs (eq.~\ref{eq:metric}) and evaluating the Ricci tensor,
\begin{equation}
R_{\mu\nu}
=
\partial_{\lambda}\Gamma^{\lambda}_{\mu\nu}
-
\partial_{\nu}\Gamma^{\lambda}_{\mu\lambda}
+
\Gamma^{\lambda}_{\mu\nu}\Gamma^{\sigma}_{\lambda\sigma}
-
\Gamma^{\sigma}_{\mu\lambda}\Gamma^{\lambda}_{\nu\sigma},
\end{equation}
where
\begin{equation}
\Gamma^{\lambda}_{\mu\nu}
=
\frac{1}{2}g^{\lambda\sigma}
\left(
\partial_{\mu}g_{\nu\sigma}
+
\partial_{\nu}g_{\mu\sigma}
-
\partial_{\sigma}g_{\mu\nu}
\right).
\end{equation}
Since this is spherically symmetric system, we can compute the Ricci tensor at fixed angular points without losing generality. In this case it is evaluated on the equatorial slice $\theta=\pi/2$.
We then minimize the Einstein field equations residuals in vacuum over the collocation points by minimizing the physics loss,
\begin{equation}
\mathcal{L}_{\rm phy}
=
\frac{1}{N_{\rm col}}
\sum_{j=1}^{N_{\rm col}}
\frac{1}{16}
\sum_{\mu,\nu=0}^{3}
\left[
R_{\mu\nu}(t_j,r_j)
\right]^2 .
\end{equation}

For the forecast runs (Sec.~\ref{sec:results.cauchy}), we also include gauge conditions in the PDE loss, namely the $1+\log$ slicing condition and the Gamma-driver shift condition. These are the same gauge conditions used to produce reference \ac{nr} solution \cite{Bernuzzi:2009ex}. The vacuum Einstein equations $R_{\mu\nu}=0$ are covariant, so they do not depend on the choice of coordinates. Enforcing them alone does not determine how the time and spatial coordinates evolve in the region where no data is provided. Including the  gauge conditions removes this ambiguity.

The lapse residual is given as ,
\begin{equation}
\mathcal{G}_{\rm lapse}
=
\partial_t \alpha
-
\beta^r \partial_r \alpha
+
2\alpha K,\label{eq:lapse}
\end{equation}
where $K=\gamma^{ij}K_{ij}$ is the trace of the extrinsic curvature $K_{ij}$. The
corresponding loss is
\begin{equation}
\mathcal{L}_{\rm lapse}
=
\frac{1}{N_{\rm col}}
\sum_{j=1}^{N_{\rm col}}
\left[
\mathcal{G}_{\rm lapse}(t_j,r_j)
\right]^2.
\end{equation}
For the shift, we use the Gamma-driver residual
\begin{equation}
\mathcal{G}_{\rm shift}
=
\partial_t \beta^r
-
\beta^r \partial_r \beta^r
+
\eta \beta^r
-
\mu_S \tilde{\Lambda}^{r},\label{eq:shift}
\end{equation}
with $\eta=2$, $\mu_S=1$ and $\tilde{\Lambda}^{r}
=
\frac{\partial_r \tilde{\gamma}_{rr}}{\tilde{\gamma}_{rr}^{2}}
-
\frac{2}{r\tilde{\gamma}_{rr}}
+
\frac{2\sqrt{\tilde{\gamma}_{rr}}}{r}$.
The corresponding loss is
\begin{equation}
\mathcal{L}_{\rm shift}
=
\frac{1}{N_{\rm col}}
\sum_{j=1}^{N_{\rm col}}
\left[
\mathcal{G}_{\rm shift}(t_j,r_j)
\right]^2 .
\end{equation} 

All the derivatives for calculating the PDE residuals are computed using fourth order finite differencing in the physical coordinates (t,r). We fix the stencil spacing to $h_r=h_t = 0.01$ for the entire $(t,r)$ domain.

\subsection{Network Architecture}
Our network architecture uses a separable physics-informed neural network (SPINN) \cite{spinn}. The solution is represented by two independent 1-D sub-networks, one for each input coordinate, whose outputs are combined to form the full solution. Each sub-network takes a single normalized input coordinate \(c\) and consists of four hidden layers of width 128 with \(\tanh\) activations. We adopt the modified multilayer perceptron (MLP) given by Wang \textit{et al.}~\cite{mlpmod} in which each hidden layer has access to features computed from the original input. This architecture has been shown to improve gradient flow pathologies in \ac{pinn} training as shown in \cite{mlpmod}. 

Specifically, the input is first mapped to three feature vectors:
\begin{align}
U &= \tanh(W_Uc+b_U), &\\
V &= \tanh(W_Vc+b_V), &\\
H^{(0)} &= \tanh(W_0c+b_0), &
\end{align}
where $W_{(\cdot)}$ and $b_{(\cdot)}$ are learnable weights and biases. The vectors \(U\) and \(V\) are shared across all subsequent hidden layers. Starting
from the first layer \(H^{(0)}\), each of the next three layers compute a mixing vector
\(Z^{(\ell)}\) and use it to update the output by:
\begin{align}
Z^{(\ell)}
    &= \tanh\!\left(W_\ell H^{(\ell-1)}+b_\ell\right),&\\
H^{(\ell)}
    &= \left(1-Z^{(\ell)}\right)\odot U
       +Z^{(\ell)}\odot V,&
\end{align}
where $\ell=1,2,3$ and \(\odot\) denotes element-wise multiplication. Finally, a linear layer produces the
sub-network output:
\begin{equation}
F(c)=W_{\mathrm{out}}H^{(3)}+b_{\mathrm{out}}
    \in\mathbb{R}^{4K}.
\end{equation}

The time sub-network outputs $F_t(\hat t) \in \mathbb{R}^{4K}$ and the radial
sub-network outputs $F_r(\hat r) \in \mathbb{R}^{4K}$. These outputs are reshaped
into $4 \times K$ feature arrays with entries $T_{d,k}(\hat t)$ and
$R_{d,k}(\hat r)$, where $d = 1, \dots, 4$ indexes the output fields and
$k = 1, \dots, K$ indexes the rank. Each field is then reconstructed as a
rank-$K$ sum of separable products plus a learnable per-field offset:
\begin{equation}
  z_d(\hat t, \hat r) = \sum_{k=1}^{K} T_{d,k}(\hat t)\, R_{d,k}(\hat r) + \beta_d,
  \label{eq:spinn}
\end{equation}
where we use $K = 64$.

All weights of the two sub-networks are initialized with Glorot normal initialization \cite{pmlr-v9-glorot10a} and biases are set to zero. The trainable output bias $b_d$ is initialized to the mean of field $d$ over the labeled training data. Finally, we scale the final-layer weights of each sub-network by $10^{-2}$ at initialization, so that the initial prediction is close to $b_d$.

\subsection{Training}
\label{sec:methods.training}

We train the network by minimizing the loss in Eq.~\eqref{eq:loss}. The full physics loss, including the Ricci and gauge terms, is
multiplied by a factor that rises linearly from \(0.1\) in the first epoch to
\(1\) by the tenth epoch. This introduces the physics constraints gradually
while the network begins to fit the labeled data. Beyond this warm-up, the region where physics is enforced depends on the application. For the interpolation problem discussed in sec.~\ref{sec:results.interp}, labels cover the whole training domain, so collocation points are drawn everywhere from the beginning of training. For the Cauchy and BVP problems in sec.~\ref{sec:results.cauchy}, and sec.~\ref{sec:results.bval}, the
collocation points are initially restricted to the neighborhood of the labeled region, and this region is gradually enlarged during the training, specifically forward in time for the Cauchy problem and inward in radius for the boundary value problem. This curriculum training \cite{krishnapriyan2021characterizingpossiblefailuremodes} helps the network to first satisfy the physics equations where the data constrains the solution well, and then carry that solution step by step into the region without labeled data.

Optimization is performed in two stages. We first train the network with the AdamW optimizer \cite{adamw} using mini-batches of labeled data and collocation points. The initial learning rate is set to \(10^{-3}\), with decoupled weight decay \(10^{-4}\) and a cosine learning-rate schedule. We train for 1000 epochs of 200 steps each. At each optimization step we draw a minibatch of 8192 labeled data points and 2048 collocation points at which the physics residuals are evaluated. The collocation points are sampled uniformly within two regions: half over the full radial domain and half in the near-puncture region $r\le15$ (see Sec.~\ref{sec:methods.input} for full domain details). This choice of $r=15$ is empirical. Close to the puncture the fields vary steeply and the gauge dynamics are non-trivial, so the equations are hardest to satisfy there, and thus this region benefits from a higher density of collocation points.

For Sec.~\ref{sec:results.cauchy} and ~\ref{sec:results.bval}, we further refine the solution with the Levenberg--Marquardt (LM) method~\cite{LMoptim}, a damped Gauss--Newton scheme for nonlinear least-squares problems. To limit the computational cost of constructing the residual Jacobian, each LM step uses a subset of points---1024 labeled points and 2048 collocation points drawn from a scrambled Sobol sequence, which covers the domain more evenly than a single random draw. The points are drawn uniformly in $t$ and $\log r$, resolving
the region near the puncture. Both sets are held fixed for 15 steps and then redrawn to avoid overfitting. All residuals for these points are collected into a single vector $\mathbf \rho(\Theta)$, where $\Theta$ denotes the trainable parameters of the network; the errors include data errors of the labeled fields, and, at each
collocation point, the independent Ricci components and the lapse and shift residuals. The entries are weighted so that
\begin{equation}
  \mathcal{L}_{\rm LM}(\Theta) = \|\mathbf \rho(\Theta)\|_2^2
\end{equation}
is the same loss as in Eq.~\eqref{eq:loss}, evaluated on the fixed points. At each iteration we compute the Jacobian $J = \partial\mathbf \rho/\partial\Theta$ and propose the update
\begin{equation}
  \Theta_{\rm trial} = \Theta - J^{\mathsf T}\big(JJ^{\mathsf T} + \lambda I\big)^{-1}\mathbf \rho(\Theta).
\end{equation}
 where $\lambda>0$ is a damping parameter. A step is accepted only if it lowers $\mathcal{L}_{\rm LM}$, after which $\lambda$ is decreased. Otherwise $\lambda$ is increased and the step is recomputed. The initial value of $\lambda$ is taken as $10^{-8}$. We perform 1500 LM iterations for each application. 

We perform the training on  a single \texttt{NVIDIA A100-40GB} GPU in double precision. Each training took approximately 8 hours on a single GPU.

\section{Results}
\label{sec:results}

\subsection{Physics-Informed Interpolation}
\label{sec:results.interp}

As first application of the method that we have developed, we consider the
construction of efficient, differentiable representations of \ac{nr} spacetimes. In
this context, the \ac{pinn} is trained using \ac{nr} data representing a finite region of
a spacetime of interest and can then be used as a more efficient representation
of the underlining \ac{nr} data for further study, e.g., to identify apparent
horizons, or compute geodesics. An advantage of this approach is that, because
the \ac{pinn} loss also encodes the Einstein field equations, they can be robustly
used to interpolate and even extrapolate the training data reliably. The utility
of \acp{pinn} in this context have already been demonstrated by
\cite{Cranganore:2025vsu}. The main difference in our work is that our
architecture is trained to predict the conformal factor $\chi$ and the conformal
metric $\tilde\gamma_{rr}$, both of which remain regular even at the center of
punctures. Instead \cite{Cranganore:2025vsu} generate the spacetime metric
directly.

For this test, the \ac{nr} data is obtained
using the 1D (spherically symmetric) Z4c implementation described in
Ref.~\cite{Bernuzzi:2009ex}. The \ac{nr} simulation describes a single, nonrotating
BH with $\Delta r = 0.01$ and $\mathrm{CFL} = 0.5$. We selected ${N_{\rm tot}} = 3.4 \times 10^{6}$ spacetime coordinates on 201 uniformly spaced time slices, spanning $t \in [0, 100]$ and $r \in [0.505, 170]$. At each time slice, alternate radial grid points were assigned to the held-out test set, while the remaining radial points were used for training. We fix the weights in the loss function as $w_{\rm data} = 10^3, w_{\rm phy} =1, w_{\rm lapse} = 0$ and $w_{\rm shift}=0$. We do not explicitly enforce gauge conditions through the PDEs in this application.

\begin{figure*}[!t]
    \includegraphics[width=0.8\textwidth]{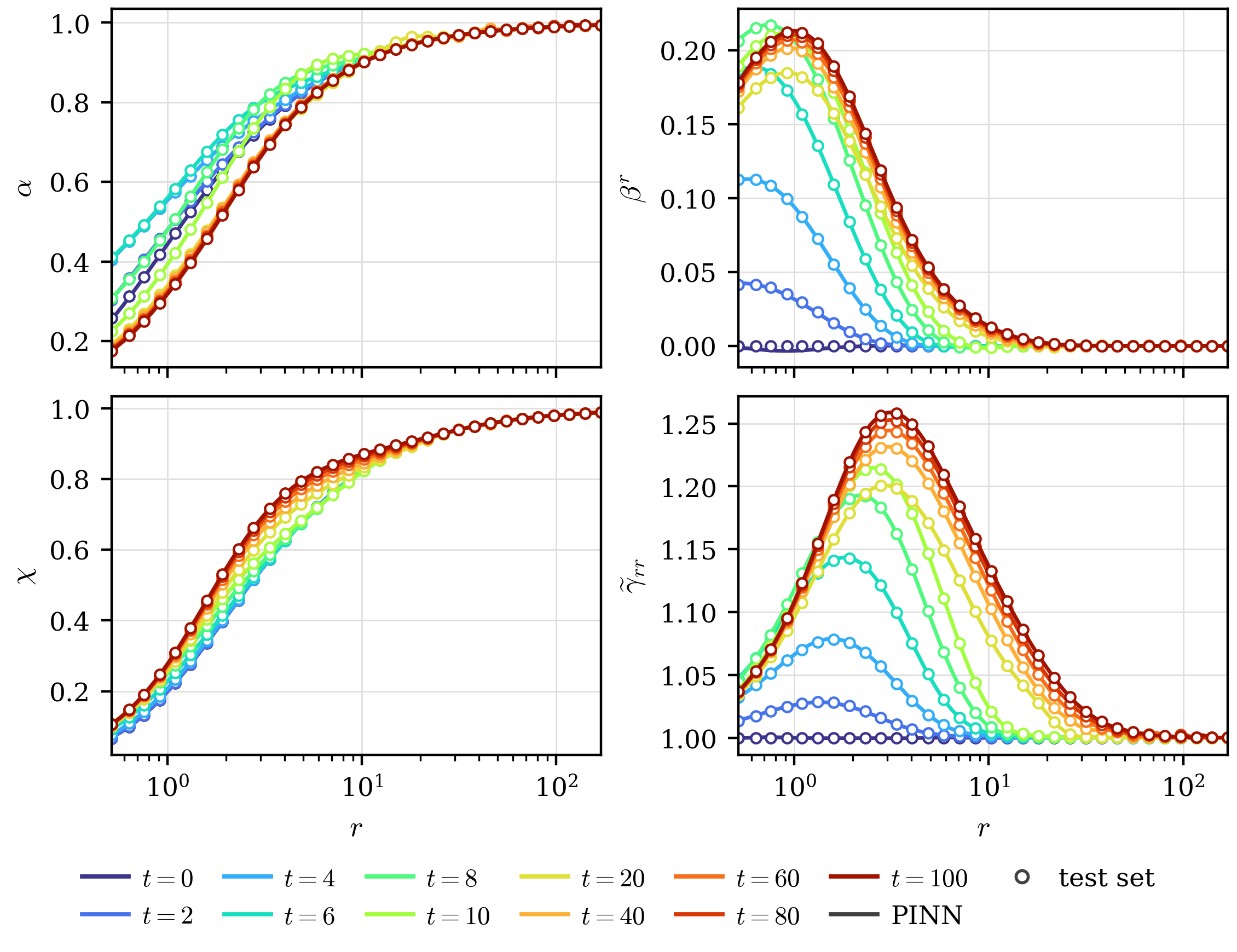}
    \caption{\ac{pinn} interpolation of the \ac{nr} data (Sec~\ref{sec:results.interp}). \ac{pinn} reconstruction (lines) of the four evolved fields compared with held-out \ac{nr} test data (circles) are shown at selected times. The network is trained on the entire \ac{nr} domain, $t\in[0,100]$ and $r\in[0.505,170]$, with the data loss and the physics loss enforced everywhere; the test set consists of grid points that are not used in training. The early times ($t\le10$) show the rapid gauge evolution from the pre-collapsed lapse towards the trumpet solution. The \ac{pinn} reproduces this evolution and is able to accurately represent the \ac{nr} data.}
    \label{fig:interp.profiles}
\end{figure*}

Our results are shown in Fig.~\ref{fig:interp.profiles}. The initial data
corresponds to a puncture with pre-collapsed lapse function $\alpha$. As time
progresses, the puncture gauge drives the system towards a trumpet solution.
This process is documented in detail in Refs.~\cite{Hannam:2008sg,
Brown:2007tb}, to which we refer for further details. Here, we note that the
coordinate location of the horizon quickly expands from $r = 0.5$ at $t
= 0$ to $r \simeq 1$ at $t=7$ and then continues to slowly expand. By $t=100$, the shift vector peaks at $\beta^r \simeq 0.2$ and the conformal
metric peaks at $\tilde\gamma_{rr} \simeq 1.25$ . The \ac{pinn} is able to correctly
encode this evolution, including in the strong field region within the BH
horizon, where the physical metric becomes very large (since
$\gamma_{rr} = \tilde\gamma_{rr}/\chi$ and $\chi \to 0$ near the singularity).

\begin{figure}
    \includegraphics[width=\columnwidth]{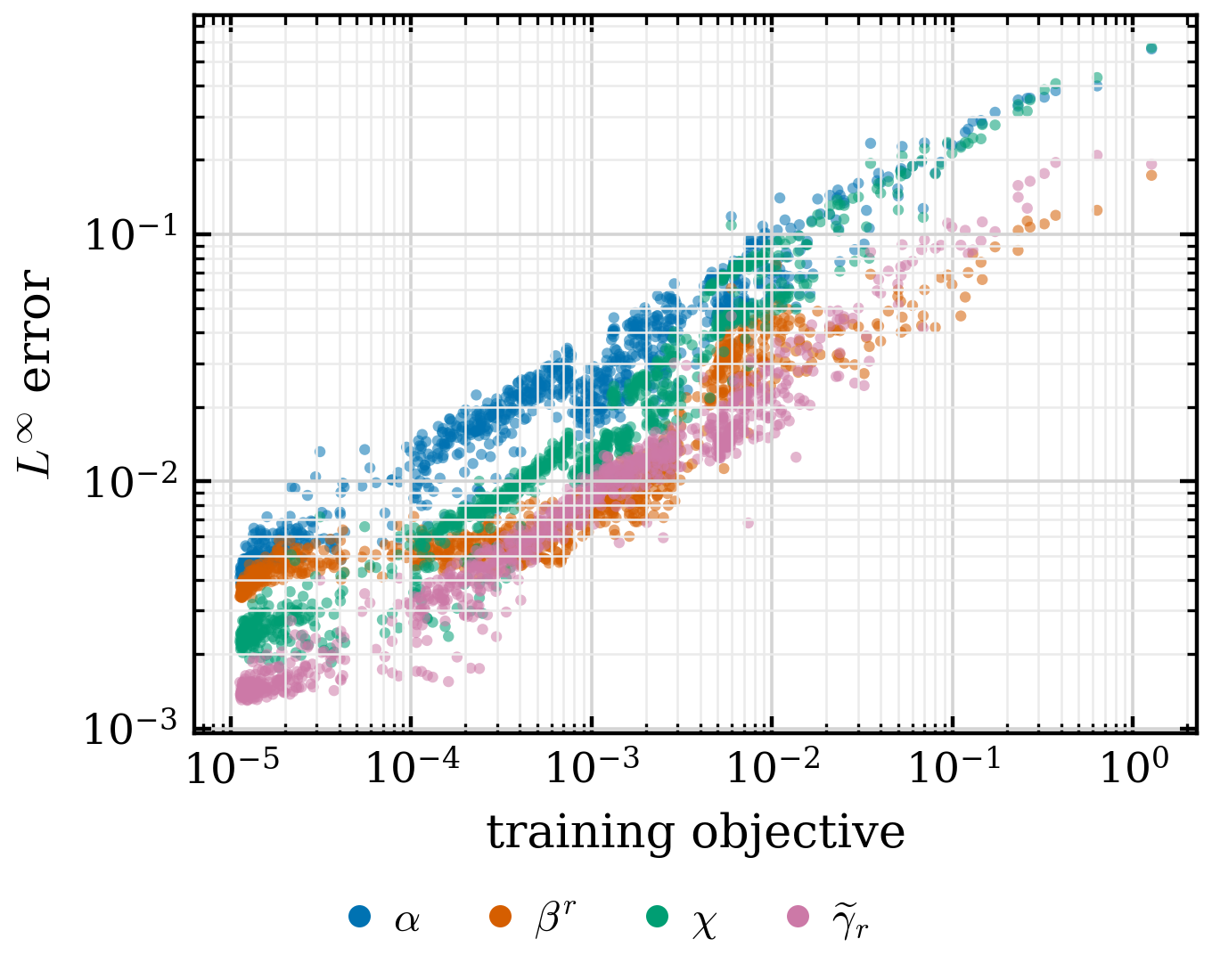}
    \caption{$L^\infty$-error for the \ac{pinn}-reconstruction as a function of the training loss error. The two error measures are well correlated over the region of interest for the loss function. The training objective, which is available without access to the ground truth, is therefore a reliable proxy for the true error of the reconstruction.}
    \label{fig:interp.loss}
\end{figure}

For this particular application, we are evaluating the \ac{pinn} at locations at
which the ``ground truth'' solution from \ac{nr} is available. As such, we are able
to compute, e.g., the norm of the error in each quantity of interest. However,
in general, only the loss function (Eq.~\ref{eq:loss}) is available to us. We
wish to assess whether the latter is a ``good'' error norm. The results are
shown in Fig.~\ref{fig:interp.loss} where we compare the $L^\infty$-norm of the
error in each of the fields predicted by the \ac{pinn} with the nominal value of the
loss function for the same field. As we can see, the two measures are strongly
correlated down to losses of $\mathcal{O}(10^{-3}){-}\mathcal{O}(10^{-2})$.
Subsequent improvements in the loss function result in minor or no improvement
in the $L^\infty$-error for $\tilde{\gamma}_{rr}$, because the loss function
becomes dominated by derivative terms. In other words, as the loss is reduced
below this threshold value, the \ac{pinn} improves in its satisfaction of the
Einstein equations, but not in its agreement with the \ac{nr} data. This is because
the \ac{nr} data is itself subject to numerical errors and does not exactly satisfy
Einstein's equations \cite{data_noise_pinn}. This tension between \ac{nr}/data-loss
and physics loss is a common theme in many of the numerical experiments we have
performed. 

\subsection{Cauchy Problem}
\label{sec:results.cauchy}

As an extension to the numerical test described in the previous section, we
consider the ability of the \ac{pinn} to evolve \ac{nr} data forward in time,
effectively solving the Cauchy problem for Einstein equations. We consider the
same \ac{nr} data described in the previous section. However, we now include in the
loss function only data with $t \leq 10$ and $r \geq 150$, effectively providing
initial data and boundary conditions to the \ac{pinn}. The physics loss is imposed in
the entire domain upto $t=60$. We fix the weights in the loss function as
$w_{\rm data} = 10^3, w_{\rm phy} = 1, w_{\rm lapse} = 1$ and  $w_{\rm
shift}=1$. Note that, in addition to the Einstein equations, the physics loss
also includes the gauge fixing conditions discussed earlier
Eqs.~\eqref{eq:lapse} and \eqref{eq:shift}. This is necessary to ensure that the
\ac{pinn} solution remains in the same gauge as the \ac{nr} data. We perform the training
using time curriculum and optimizers described in Sec.~\ref{sec:methods.training}.
During the first 150 of the 1000 AdamW epochs, the collocation points are drawn
only from $0 \leq t \leq t_{\rm hi}$, where $t_{\rm hi}$ increases linearly from
the end of the labeled interval, $t = 10$, to the final time, $t = 60$. For the
remaining epochs, and throughout the LM stage, the collocation points cover the
full interval $0 \leq t \leq 60$.

\begin{figure*}[!t]
    \includegraphics[width=0.8\textwidth]{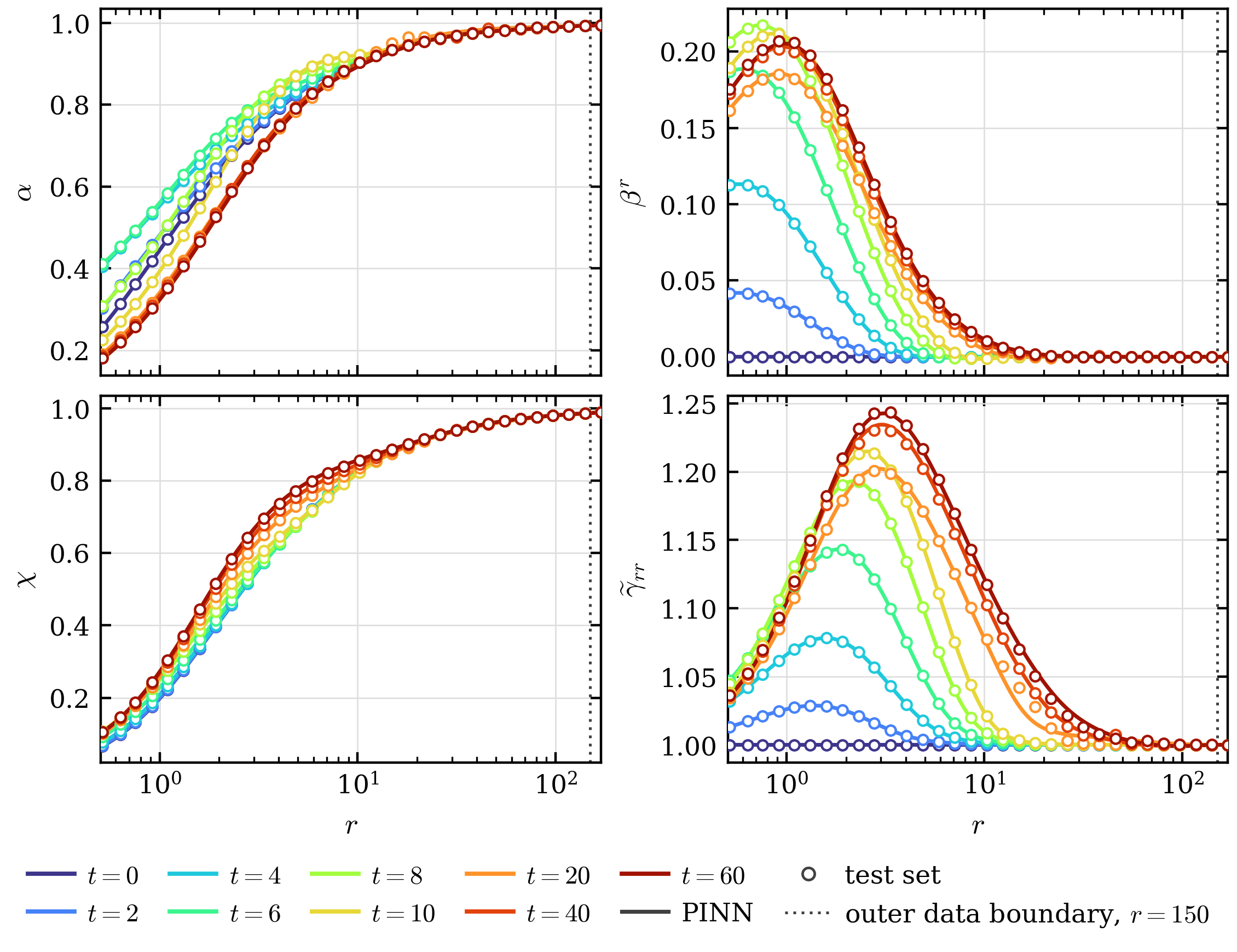}
    \caption{\ac{pinn} evolution of \ac{nr} data (Sec~\ref{sec:results.cauchy}). \ac{pinn} reconstruction (lines) of the four evolved fields compared with held-out \ac{nr} test data (circles) are shown at selected times. The \ac{pinn} is trained with labeled data only for $t\le10$ (initial data) and $r\ge150$ (outer boundary, dotted line), and solves the Einstein equations together with the 1+log and Gamma-driver gauge conditions on the full domain $t\in[0,60]$, $r\in[0.505,170]$. The profiles at $t>10$ are therefore a forecast with no labeled data in the plotted region $r<150$. The \ac{pinn} carries the puncture through the transition to the trumpet solution and up to $t=60$. In the forecast region it reproduces the qualitative features of the evolution: the lapse continues to collapse at the puncture and then settles, the shift grows and develops its peak near the horizon, and the peak of $\tilde\gamma_{rr}$ grows and moves outward before settling to the stationary trumpet profile. The \ac{pinn} and \ac{nr} curves are nearly indistinguishable for $\alpha$, $\beta^r$ and $\chi$. The visible deviations in $\tilde\gamma_{rr}$ at $r\sim20$, are attributed to a small outgoing gauge wave that \ac{pinn} is unable to capture (see Fig.~\ref{fig:cauchy.gauge} and related discussion).}
    \label{fig:cauchy.profiles}
\end{figure*}
\begin{figure*}
    \includegraphics[width=0.8\textwidth]{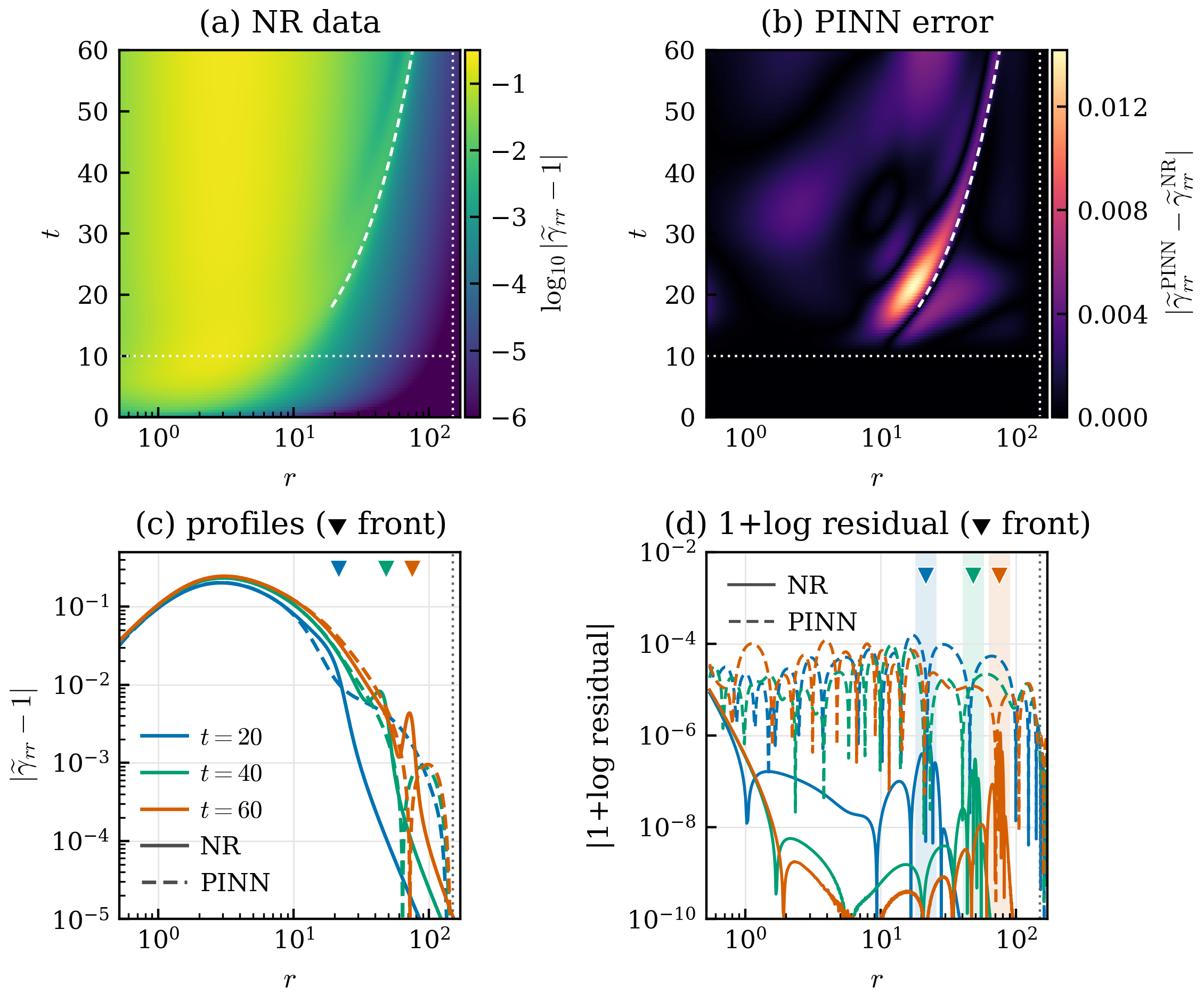}
    \caption{An outgoing gauge wave is visible in the conformal metric field that the \ac{pinn} does not resolve. (a) \ac{nr} data, $\log_{10}|\tilde\gamma_{rr}-1|$, over $(t, r)$. The main peak of $\tilde\gamma_{rr}$ settles at $r \approx 3$, while a small, gauge wavefront (dashed) starts from  $t \approx 18$ and travels outward, from $r \approx 19$ to $r \approx 76$ by $t = 60$. Dotted lines mark the labeled regions, $t \leq 10$ and $r \geq 150$, indicating that the front propagates entirely through the unlabeled forecast region where only physics is enforced. (b) \ac{pinn} error $|\tilde\gamma_{rr}^{\rm PINN}-\tilde\gamma_{rr}^{\rm NR}|$: the
largest errors lie in a band along this wavefront. (c) Profiles of $|\tilde\gamma_{rr}-1|$ at $t = 20, 40$ and $60$ (\ac{nr} solid, \ac{pinn} dashed; triangles mark the moving wavefront): the \ac{pinn} is able to capture the main features but misses the small gauge wave resulting in poor approximation in that region. (d) Residual of the 1+log slicing condition at the same times, computed with the fourth-order finite differencing for both \ac{nr} data and \ac{pinn} prediction (shading and triangles mark the wavefront). The \ac{nr} data satisfy the condition across the front to $\mathcal{O}( 10^{-7})$ with a visible bump where the gauge wavefront moves. The \ac{pinn} residual is $\mathcal{O}(10^{-5})$ throughout, which is two to four orders of magnitude larger and barely rises at the front: at this level the physics loss does not capture the wavefront, which is smeared out in the \ac{pinn} solution}.

    \label{fig:cauchy.gauge}
\end{figure*}

\begin{figure*}
    \includegraphics[width=\textwidth]{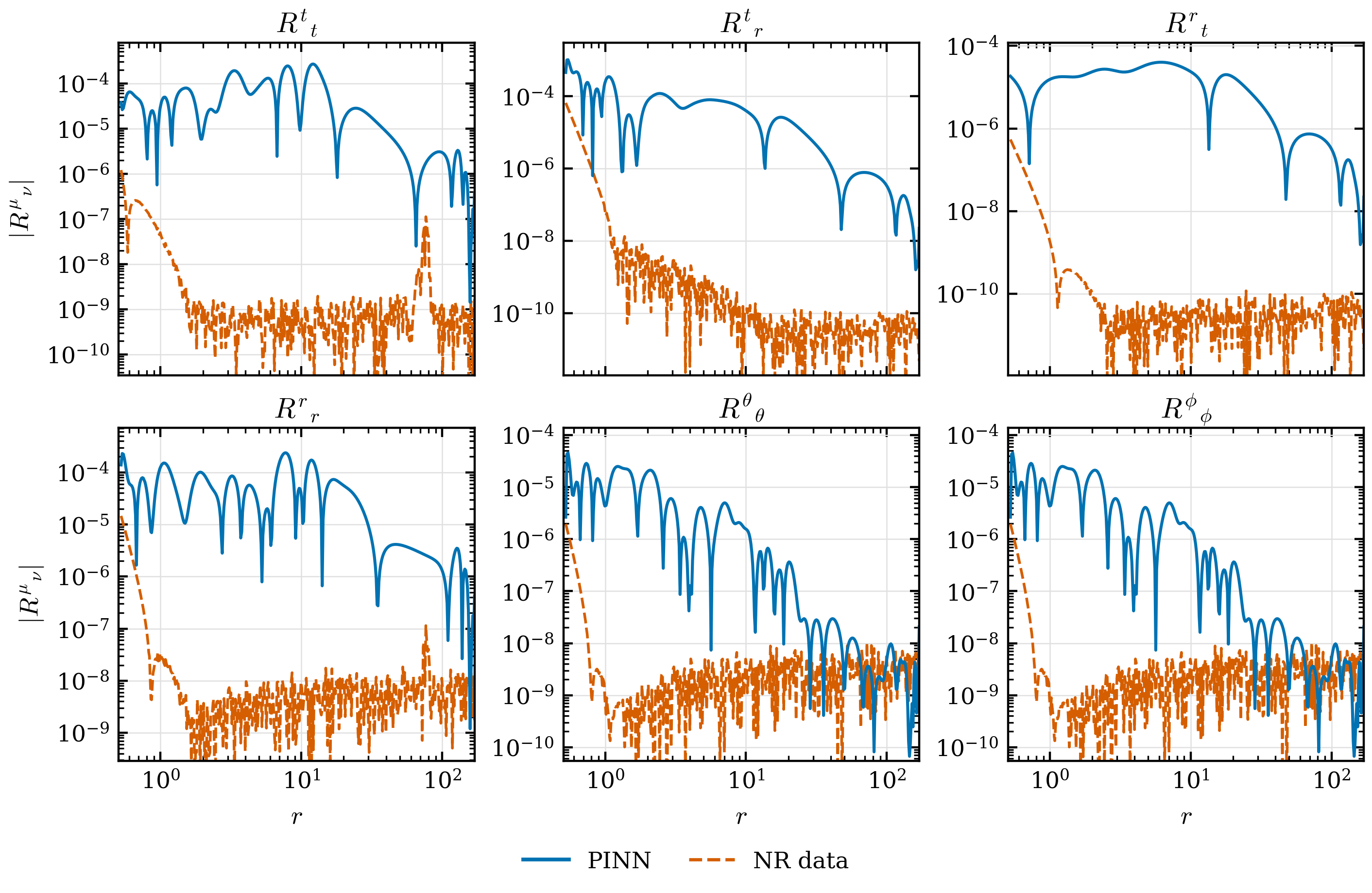}
    \caption{Mixed-index Ricci components \(|R^{\mu}{}_{\nu}|\) at \(t=60\,M\) as a function of \(r\). The \ac{pinn} residual (solid blue) is computed from the network output while the \ac{nr} residual (dashed orange) is computed using the \ac{nr} data. Since the solution is vacuum, the exact value is zero for all components. The \ac{nr} curve therefore indicates the truncation-error and noise floor of the reference data, about \(\mathcal{O}(10^{-9})\). The \ac{pinn} residual is \(\mathcal{O}(10^{-5})\) for \(R^t{}_t\), \(R^t{}_r\), \(R^r{}_t\) and \(R^r{}_r\) over most of the domain, and decreases to \(\mathcal{O}(10^{-8})\) at large \(r\) for \(R^\theta{}_\theta\) and \(R^\phi{}_\phi\). Sharp dips correspond to sign changes of the residual.}
    \label{fig:cauchy.ricci}
\end{figure*}
Figure \ref{fig:cauchy.profiles} shows the result of our numerical experiment.
We see that the \ac{pinn} is able to capture the qualitative evolution of the data at later times, but error persists most visibly in the conformal metric $\tilde\gamma_{rr}$ around later times at $r \sim 20$. This error is associated with a small, sharp gauge wavefront that propagates outward through the unlabeled region (see Fig. \ref{fig:cauchy.gauge}~c); the \ac{pinn} reproduces the main features of the solution but is unable to capture the smaller features such as this wavefront. As shown in Fig. \ref{fig:cauchy.gauge}~d, at the residual level in the gauge conditions that the \ac{pinn} attains, the gauge loss is unable to capture the smaller features of the solution. Overall, however, the \ac{pinn} agrees well with the \ac{nr} data in the forecast region where no physics data is shown.

The mixed Ricci components (Fig.~\ref{fig:cauchy.ricci}) quantify how well the \ac{pinn} satisfies the field equations. At $t = 60$ the magnitude of \ac{pinn} residual is $ \mathcal{O}(10^{-5})$ for the $(t, r)$ components, three to four orders of magnitude higher than the \ac{nr} data, whose own components computed with the same finite-difference operator show $ \mathcal{O}(10^{-9})$ error. Since the \ac{nr} solution satisfies both the labeled data and the field equations to this accuracy, the gap reflects the finite accuracy reached by the network and its optimization, a known limitation of \ac{pinn} training \cite{mlpmod}. We remark that even though the \ac{pinn} error is larger compared to \ac{nr}, it reaches sufficiently low values to capture to ground truth both qualitatively and quantitatively.

\subsection{Boundary Values Problem}
\label{sec:results.bval}
As a final test, we consider the ability of \acp{pinn} to solve inverse problems for the Einstein equations. We generate analytical data for this application and prescribe the spacetime metric for $r \geq 15$ to be that of a unit-mass Schwarzschild BH in Kerr-Schild coordinates (see Appendix~\ref{app:ks} for more details). We test the ability of the \ac{pinn} to reconstruct the interior solution in $2 \leq r < 15$, down to the horizon location $r=2$ in Kerr-Schild coordinates. Specifically, our loss function includes the data loss term only for $r \geq 15$, while the physics loss is imposed upto the radius $r=2$. As it is well known,
the vacuum solution of Einstein equations in spherical symmetry is unique (Birkhoff's theorem). However, because we do not prescribe any condition for the gauge fields, the problem we are considering is \emph{ill posed} and admits
infinite solutions. As such this test case allows us to explore the use of \acp{pinn} to solve generic inverse problems for Einstein equations, which are typically ill posed.

\begin{figure*}
\includegraphics[width=\textwidth]{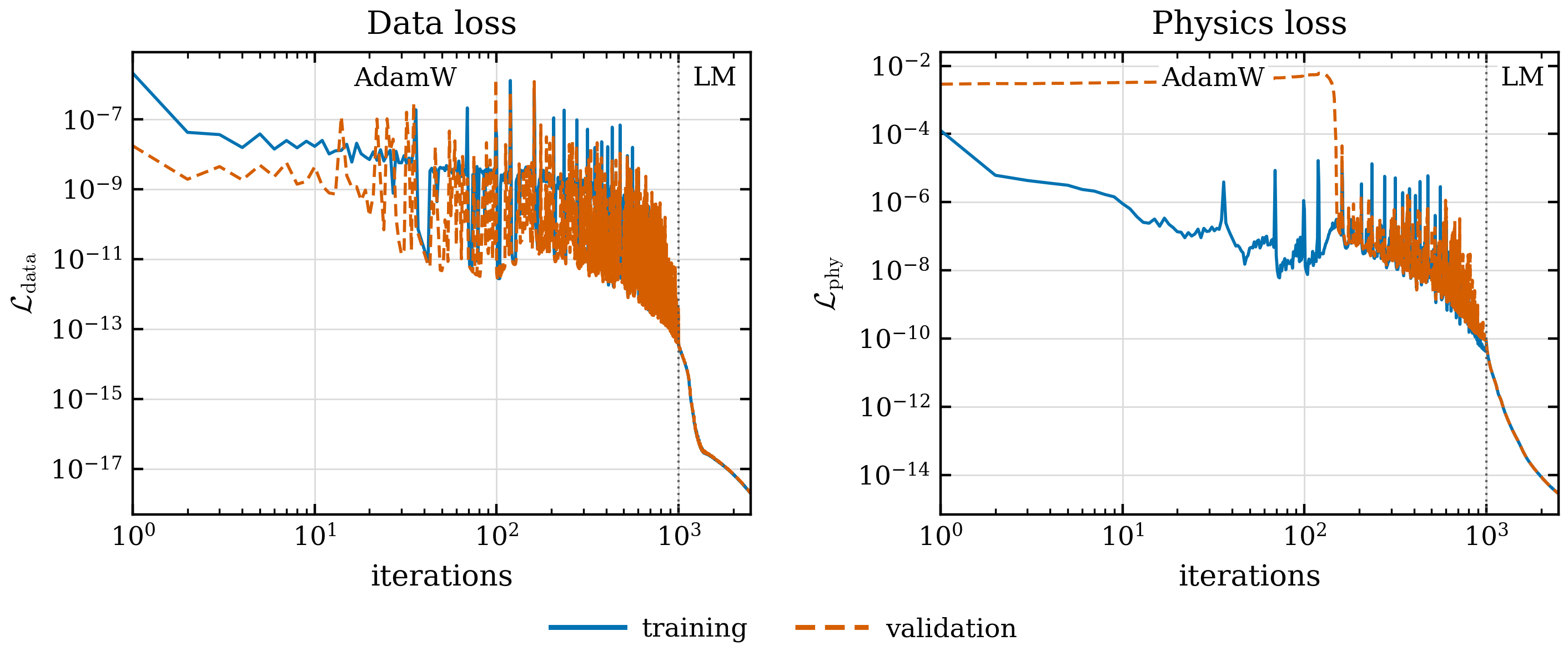}
    \caption{Data loss (left) and physics loss (right) for the boundary-value problem, on the training points (solid) and on held-out validation points (dashed). Training uses 1000 AdamW epochs followed by 1500 Levenberg--Marquardt (LM) iterations (dotted line). The validation physics loss, evaluated over the whole domain, stays at $\sim3\times10^{-3}$ until epoch $\approx150$, when the radial curriculum has moved the physics constraint from $r=15$ in to the horizon at $r=2$; it then drops to the training value. AdamW stalls with large oscillations at $\mathcal{L}_{\rm data}\sim10^{-13}$ and $\mathcal{L}_{\rm phy}\sim10^{-10}$. LM lowers both by a further four to five orders of magnitude, to $\mathcal{L}_{\rm data}\sim2\times10^{-18}$ and $\mathcal{L}_{\rm phy}\sim3\times10^{-15}$, with training and validation curves following same trend (no overfitting). The second-order LM stage is what drives the solution to low values; these values are reachable because the Kerr--Schild data are analytic and noise-free.}
    \label{fig:bc.loss}
\end{figure*}

We fix the weights of the loss function (eq.~\ref{eq:loss}) to $w_{\rm data} = 10^3, w_{\rm phy} =1, w_{\rm lapse} = 0$ and $w_{\rm shift}=0$. As mentioned previously, we do not fix the gauge in this application. During the first 150 epochs we use a radial curriculum: the inner boundary of the collocation region moves linearly from $r = 15$ to the horizon at $r = 2$, so the physics constraint is extended inward gradually from the data region. We follow the training methodology described in sec.~\ref{sec:methods.training}. As demonstrated in Fig.~\ref{fig:bc.loss}, stage 1 denotes the training with AdamW optimizer and stage 2 with LM optimizer. Both mean squared error (MSE) for validation (dashed lines) and training (solid) dataset reduce with each epoch. The data loss is computed for labeled data with $r \geq 15$ and the physics loss is computed on the entire domain. The errors go down to low values $\mathcal{O}(10^{-16})$ because the training data is noise-free and analytically generated. In the previous case (see sec.~\ref{sec:results.cauchy}), it was harder for the optimizers to minimize due to numerical noise in the data which doesn't fully satisfy the PDEs embedded in the loss function.

\begin{figure*}
    \includegraphics[width=0.8\textwidth]{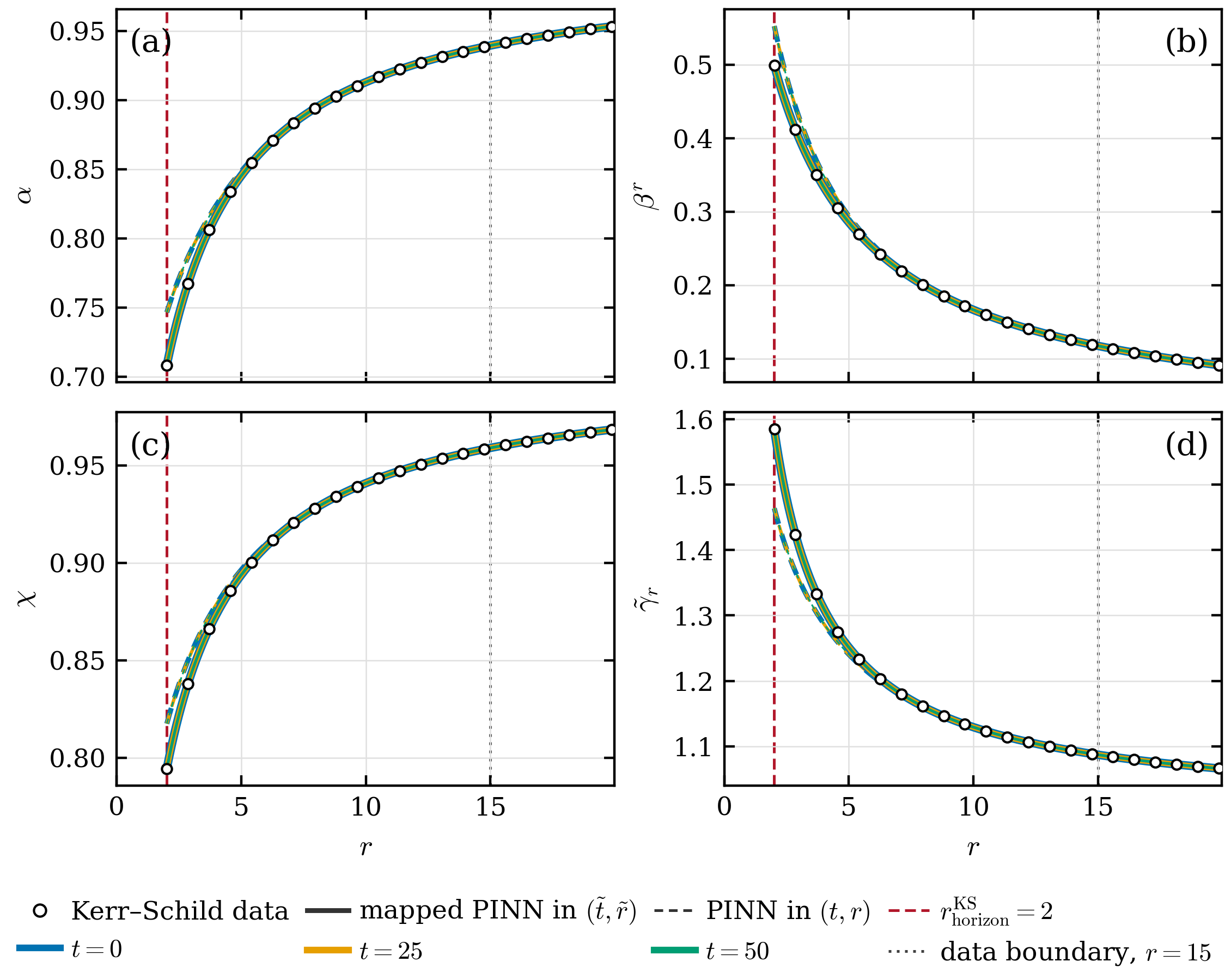}
    \caption{Reconstruction of the Schwarzschild interior from exterior data (Sec~\ref{sec:results.bval}). The \ac{pinn} is trained on $t\in[0,50]$, $2\le r\le20$, with Kerr--Schild data only shown at $r\ge15$ (dotted line) and with only the vacuum Einstein equations, without gauge conditions, in $2\le r<15$. Panels (a)--(d) show $\alpha$, $\beta^r$, $\chi$ and $\tilde\gamma_{rr}$. Circles are the analytic Kerr--Schild fields, dashed lines are the \ac{pinn} in its own coordinates $(t,r)$, and solid lines the \ac{pinn} after the learned coordinate map to $(\tilde t,\tilde r)$, at $t=0$, 25 and 50 (colors). Inside $r<15$ the unmapped \ac{pinn} differs from Kerr--Schild by up to $11\%$ (in $\beta^r$ at $r=2$) because it has converged to a different gauge. After the coordinate transformation to Kerr--Schild, all four fields agree with ground truth to better than $8\times10^{-4}$ (relative) down to the horizon (red dashed line). The three time slices lie on top of each other: although stationarity is not imposed, the \ac{pinn} fields change by less than $2\times10^{-4}$ (relative) between $t=0$ and $t=50$. The \ac{pinn} therefore recovers the Schwarzschild solution, as required by Birkhoff's theorem, in coordinates of its own choosing.}

    \label{fig:bc.fields}
\end{figure*}
Figure~\ref{fig:bc.fields} shows the reconstructed solution from the \ac{pinn} method
(dashed lines), as well as the analytic solution (dots). Interestingly, the two
disagree in the interior region ($r < 15$), because the \ac{pinn} converges to a solution in a different gauge from that of the Kerr-Schild
coordinates. In other words, while the data is plotted as a function of $r$ for
both data sets---the \ac{pinn} data and the analytic solution)---the meaning of $r$
differs in the two cases. To be able to properly compare the two, we implement
the following procedure to discover the coordinate transformation connecting the
\ac{pinn} spacetime solution to Kerr-Schild coordinates.

We train a second NN,
$\mathbf{x}' = \Phi(\mathbf{x})$, that maps the \ac{pinn} coordinates
$\mathbf{x} = (t, r, \theta, \phi)$ to new coordinates
$\mathbf{x}' = (t', r', \theta', \phi')$ in which the metric should take the
known Kerr--Schild form. A metric transforms under a change of coordinates as
\begin{equation}
    g'_{\mu\nu}(\mathbf{x}') = \frac{\partial x^{\alpha}}{\partial x'^{\mu}}
    \frac{\partial x^{\beta}}{\partial x'^{\nu}}\, g_{\alpha\beta}(\mathbf{x}),
\end{equation}
so we train $\Phi$ by minimizing the loss function , 
\begin{equation}
    \mathcal{L}_{\rm map} = \frac{1}{16N}\sum_{i=1}^{N}\sum_{\mu,\nu=0}^{3}
    \left[ g^{\rm KS}_{\mu\nu}(\mathbf{x}'_i)
    - \frac{\partial x^{\alpha}}{\partial x'^{\mu}}
      \frac{\partial x^{\beta}}{\partial x'^{\nu}}\,
      g^{\rm PINN}_{\alpha\beta}(\mathbf{x}_i) \right]^2 ,
\end{equation}
where $g^{\rm PINN}_{\alpha\beta}$ is the metric assembled from the \ac{pinn} output and
$g^{\rm KS}_{\mu\nu}$ is the analytic Schwarzschild metric in Kerr--Schild
coordinates \cite{hiddensym}. The Jacobian
$\partial x'/\partial x$ is obtained from $\Phi$ by automatic differentiation and
inverted at each point. Since both metrics are spherically symmetric, we place all
training points on the equatorial plane without loss of generality, specifically $\theta = \pi/2$ and $\phi = 0$. If the coordinate transformation discovered by minimizing this loss allows use to map the transformed fields to the known analytical values, we can conclude that the \ac{pinn} solution is the Kerr--Schild solution but written in different coordinates.

As it can be seen from Fig.~\ref{fig:bc.fields}, the mapped \ac{pinn} data (solid line) is in excellent agreement with the analytical Kerr-Schild solution, showing that the \ac{pinn} was able to successfully reconstruct the Schwarzschild metric. After the coordinate transformation, all four fields agree with the Kerr--Schild solution to a relative difference below $8\times10^{-4}$ over $t\in[0,50]$, with the largest differences near the horizon. As a final remark, we did not impose stationarity on the \ac{pinn} solution. In fact, the \ac{pinn} solution does change slightly with time. Over the full time interval $t \in [0, 50]$, the relative change of the fields, $|\Delta f/f(t=0)|$, is at most $2\times10^{-5}$ for $\alpha$, $3\times10^{-5}$ for $\beta^r$, $6\times10^{-5}$ for $\chi$, and $2\times10^{-4}$ for $\tilde\gamma_{rr}$. The largest changes occur near the horizon, far from the boundary. In the region $r \ge 15$, where data is provided, the relative change is below $\mathcal{O}(10^{-7})$.

\section{Conclusions}
\label{sec:conclusions}
In this work, we have explored the use of \acp{pinn} to solve the full set of vacuum Einstein field equations in spherical symmetry. Our approach uses a separable \ac{pinn}~\cite{spinn}: two modified MLP sub-networks~\cite{mlpmod}, one in $t$ and one in $r$, that are combined into a rank-$K$ representation of the lapse $\alpha$, the shift $\beta^r$, and the Z4c conformal variables $\chi$ and $\tilde\gamma_{rr}$. These variables stay regular at the puncture. To our knowledge, this is the first work in which a \ac{pinn} is trained on a numerical-relativity (\ac{nr}) data while enforcing the full vacuum Einstein equations along with moving-puncture gauge conditions. For various applications, we enforce curriculum training that gradually extends the region where physics is enforced into the unlabeled part of the domain where no data is available. Additionally, we use an optimization schedule in which a first-order AdamW stage is followed by a Levenberg--Marquardt refinement.

We have applied this framework to three problems of increasing difficulty. First, we used it for physics-informed interpolation of 1D Z4c puncture evolutions \cite{Bernuzzi:2009ex}. Here the \ac{PINN} is trained in the entire numerical domain spanning $t \in [0, 100]$ and $r \in [0.505, 170]$, by minimizing a loss that combines the data misfit and the physics PDE residual. We observed a good agreement between the \ac{nr} values and the \ac{pinn} predictions. We also found that the training objective is a reliable indicator of the true error (Fig.~\ref{fig:interp.loss}); as the objective falls by five orders of magnitude, the $L^\infty$ error for various output fields falls monotonically up to $\mathcal{O}(10^{-3}){-}\mathcal{O}(10^{-2})$. 

Second, we considered the Cauchy problem where the \ac{pinn} was given boundary and initial conditions. The network was trained with data only for $t \le 10$ and $r \ge 150$, and the physics and gauge conditions were enforced up to $t = 60$. The \ac{pinn} evolves the puncture forward in time and reproduces the main features of the \ac{nr} solution in the region where no data is provided. The largest discrepancies are associated with a small-amplitude gauge wave, most prominently seen in $\tilde\gamma_{rr}$, which the network does not capture. The Ricci residual is insensitive to coordinate changes and hence it does not constrain such a feature, which instead is controlled only by the gauge-condition. In this case, the 1+log loss sits two to four magnitudes larger than the \ac{nr} data error in the entire radial domain, which is consistent with the network not resolving the smaller gauge features. Finally, we observe the Ricci residuals in the components of the mixed tensor are \(\mathcal{O}(10^{-5})\) for the \ac{pinn}, compared with \(\mathcal{O}(10^{-9})\) for the \ac{nr} data. This shows that the network agrees qualitatively and quantitatively with the \ac{nr} data, although it still not as accurate.

Third, we considered an ill-posed inverse problem: reconstructing the region $2 \le r < 15$ of a Schwarzschild spacetime from analytical Kerr--Schild data at $r \ge 15$, without prescribing the gauge. Since the data is noise-free, the training and validation MSE reach low values $\mathcal{O}(10^{-16})$. We discover that the \ac{pinn} converges to a solution expressed in coordinates different from Kerr--Schild. To verify this, we trained a second NN to learn the coordinate transformation between the \ac{pinn} gauge and Kerr-Schild gauge and were able to show that the \ac{pinn} indeed converged to the expected Schwarzschild solution, as required by Birkhoff's theorem.

Overall, our results show that \acp{pinn} can evolve a moving puncture forward in time and can reconstruct a black-hole spacetime from partial data. In both cases they agree with \ac{nr} qualitatively and quantitatively. However, our results also exhibit some limitations of the current approach. Training these networks is nontrivial. To reach the accuracy reported here, we found it beneficial to combine a curriculum training that progressively extends the physics region, a higher density of collocation points near the puncture, and a damped Gauss--Newton refinement following the AdamW stage. These choices are consistent with strategies proposed in the literature to address known challenges in \ac{pinn} training
~\cite{krishnapriyan2021characterizingpossiblefailuremodes, mlpmod, rathore2024challengestrainingpinnsloss}. Addtionally, \acp{pinn} do not yet reach the accuracy of \ac{nr}. In the Cauchy problem the Ricci residual of the network is $\mathcal{O}(10^{-5})$, compared with $\mathcal{O}(10^{-9})$ for the \ac{nr} data. Reaching a given accuracy also costs more than a standard finite-difference evolution, as has been found for other partial differential equations~\cite{grossmann2023physicsinformedneuralnetworksbeat}. Lastly we observed that \acp{pinn} are sensitive to the quality of the training data. When the data carry numerical errors, as \ac{nr} data inevitably do, the data loss and the physics loss favor slightly different solutions. Beyond a certain point, further training then improves how well the Einstein equations are satisfied, but not how well the network agrees with the data~\cite{Rohrhofer_2023}. In the inverse problem of Sec.~\ref{sec:results.bval}, where the data are exact, this conflict is absent and both losses reach substantially lower values.Nonetheless, our results demonstrate that \acp{pinn} can offer capabilities that complement those of standard \ac{nr} codes. 

From the perspective of \ac{nr}, we see several directions for future work. The natural next step is an extension to 3D spacetimes without symmetry and, eventually, to binary black-hole spacetimes. In addition, the inverse-problem setting of Sec.~\ref{sec:results.bval} suggests that \acp{pinn} can be used to reconstruct regions of a spacetime that are not directly evolved, for example the interior of a spacetime from data in the asymptotic region alone, a task that is difficult for standard evolution codes. 

\begin{acknowledgments}
ES would like to thank Rossella Gamba for insightful comments and discussions. ES and AS acknowledge support from The Charles E. Kaufman Foundation of The Pittsburgh Foundation. 
SB acknowledge funding from the EU Horizon under ERC Consolidator
Grant, no. InspiReM-101043372.
This research used resources of the National Energy Research Scientific Computing Center (NERSC), a Department of Energy User Facility using NERSC award DD3-ERCAP m5373. ES and AS also acknowledge the use of computing resources from Institute for Computational and Data Sciences (ICDS), Penn State. Claude AI was used to help with coding and plotting scripts. 
\end{acknowledgments}

\appendix

\section{Z4c fields for the Kerr--Schild solution}
\label{app:ks}

In this appendix we derive the fields $(\alpha,\beta^r,\chi,\tilde\gamma_{rr})$
used as data in Sec.~\ref{sec:results.bval}. We keep the mass $M$ general; the experiments use
$M=1$.

\subsection{Kerr--Schild coordinates}

We start from the Schwarzschild metric in Schwarzschild coordinates
$(t_{\rm S}, r, \theta, \phi)$,
\begin{equation}
  ds^2 = -\left(1-\frac{2M}{r}\right)dt_{\rm S}^2
  + \left(1-\frac{2M}{r}\right)^{-1}dr^2 + r^2 d\Omega^2 ,
\end{equation}
where $d\Omega^2 = d\theta^2 + \sin^2\theta\, d\phi^2$. This form is singular at
the horizon $r = 2M$. The ingoing Kerr--Schild time $t$ is defined by
\begin{equation}
  t = t_{\rm S} + 2M \ln\left|\frac{r}{2M}-1\right| ,
  \qquad
  dt_{\rm S} = dt - \frac{2M}{r-2M}\,dr .
  \label{eq:ks.tdef}
\end{equation}
 Substituting Eq.~\eqref{eq:ks.tdef} into the metric, gives the Kerr--Schild line element
\begin{align}
  ds^2 ={}& -\left(1-\frac{2M}{r}\right)dt^2 + \frac{4M}{r}\,dt\,dr \nonumber\\
  &+ \left(1+\frac{2M}{r}\right)dr^2 + r^2 d\Omega^2 .
  \label{eq:ks.metric}
\end{align}
All metric components are smooth across $r=2M$. Because it is $t$-independent, the corresponding data are static.

\subsection{Lapse, shift and spatial metric}

Comparing Eq.~\eqref{eq:ks.metric} with the $3+1$ line element,
$g_{tt} = -\alpha^2 + \beta_r\beta^r$, $g_{tr} = \beta_r$ and
$g_{ij} = \gamma_{ij}$, we read off
\begin{equation}
  \gamma_{rr} = 1 + \frac{2M}{r}, \qquad
  \gamma_{\theta\theta} = \frac{\gamma_{\phi\phi}}{\sin^2\theta} = r^2, \qquad
  \beta_r = \frac{2M}{r}.
\end{equation}
This gives the shift,
\begin{equation}
  \beta^r = \frac{\beta_r}{\gamma_{rr}} = \frac{2M}{r+2M} .
\end{equation}
and lapse,
\begin{equation}
  \alpha^2 = \beta_r\beta^r - g_{tt}
  = \frac{r}{r+2M} .
\end{equation}

\subsection{Conformal decomposition}

We use the conformal decomposition of Sec.~\ref{sec:methods}, with
$\gamma_{ij} = \chi^{-1}\tilde\gamma_{ij}$ and
$\det\tilde\gamma_{ij} = \det\hat\gamma_{ij}$, where
$\hat\gamma_{ij} = \mathrm{diag}(1, r^2, r^2\sin^2\theta)$. This gives
\begin{equation}
  \chi = \left(\frac{\det\hat\gamma_{ij}}{\det\gamma_{ij}}\right)^{1/3}
  = \left(\frac{r^4\sin^2\theta}{\gamma_{rr}\,r^4\sin^2\theta}\right)^{1/3}
  = \gamma_{rr}^{-1/3} .
\end{equation}

The conformal metric components are then given by,
\begin{equation}
  \tilde\gamma_{rr} = \chi\gamma_{rr} = \gamma_{rr}^{2/3}, \qquad
  \tilde\gamma_{\theta\theta} = \chi r^2 = \frac{r^2}{\sqrt{\tilde\gamma_{rr}}} .
\end{equation}

\subsection{Summary of the data}

Collecting the results, the four fields used as data are
\begin{align}
  \alpha &= \left(\frac{r}{r+2M}\right)^{1/2}, &
  \beta^r &= \frac{2M}{r+2M}, \nonumber\\
  \chi &= \left(\frac{r}{r+2M}\right)^{1/3}, &
  \tilde\gamma_{rr} &= \left(\frac{r+2M}{r}\right)^{2/3}.
  \label{eq:ks.fields}
\end{align}
We use these fields to generate analytical data for Sec~\ref{sec:results.bval}.
\bibliography{references}

\end{document}